\documentclass[
 reprint,
 twocolumn,
 amsmath,
 amssymb,
 aps,
 prl,
 nofootinbib,
 tightenlines,
 nobibnotes
]{revtex4-2}

\usepackage{graphicx}
\usepackage{dcolumn}
\usepackage{bm}
\usepackage{hyperref}

\usepackage{mathrsfs,amsmath} 
\usepackage{stfloats}
\usepackage{float}
\usepackage{gensymb}
\usepackage{braket}

\usepackage{color}

\definecolor{Green}{RGB}{0,100,0}
\definecolor{Blue}{RGB}{51,153,255}
\definecolor{Red}{RGB}{151,010,010}

\begin{document}

\title{Pump-tailored alternative Bell state generation in the first-order Hermite-Gaussian basis}

\author{Zhe Kan$^1$, Andrew A. Voitiv$^1$, Patrick C. Ford$^1$}%
\author{Mark T. Lusk$^2$}%
\email[]{mlusk@mines.edu}%
\author{Mark E. Siemens$^1$}%
\email[]{msiemens@du.edu}%
\affiliation{$^1$Department of Physics and Astronomy, University of Denver, Denver, CO 80208, USA
}%
\affiliation{
$^2$Department of Physics, Colorado School of Mines, Golden, CO 80401, USA
}%

\date{\today}

\begin{abstract}
We demonstrate entangled-state swapping, within the Hermite-Gaussian basis of first-order modes, directly from the process of spontaneous parametric down-conversion within a nonlinear crystal. The method works by explicitly tailoring the spatial structure of the pump photon such that it resembles the product of the desired entangled spatial modes exiting the crystal. Importantly, the result is an entangled state of balanced HG modes, which may be beneficial in applications that depend on symmetric accumulations of geometric phase through optics or in applications of quantum sensing and imaging with azimuthal sensitivity. Furthermore, the methods are readily adaptable to other spatial mode bases.
\end{abstract}

\maketitle

Photon pairs---biphotons---generated by spontaneous parametric down conversion (SPDC) have long been utilized to generate entangled two-particle states in low-cost, accessible experimental settings \cite{Couteau2018SpontaneousDown-conversion} to make tests of fundamental aspects of quantum physics \cite{Bouwmeester1997ExperimentalTeleportation, Kim2000DelayedEraser, Proietti2019ExperimentalIndependence} and for applications in quantum imaging and information \cite{Magana-Loaiza2019QuantumInformation}. Typically, an internal degree of freedom of the biphotons is entangled, such as polarization \cite{Kwiat1995NewPairs} or orbital angular momentum (OAM) \cite{Mair2001EntanglementPhotons}. The choice of the entangled degrees of freedom---and their associated measurement apparatus (e.g., linear polarizers)---determines which bases are suitable to describe the quantum-entangled state.

Particular interest is placed upon the Bell states since they comprise a maximally-entangled, mutually orthonormal basis for all two-photons states. The first two are defined as
\begin{subequations}
\begin{align} 
|\Phi^{+} \rangle & = \frac{1}{\sqrt{2}} \left(|0 \rangle_{\mathrm{A}} |0 \rangle_{\mathrm{B}} + |1 \rangle_{\mathrm{A}} |1 \rangle_{\mathrm{B}} \right), \label{phiplus} \\
|\Psi^{+} \rangle & = \frac{1}{\sqrt{2}} \left(\ket{0}_{\mathrm{A}}\ket{1}_{\mathrm{B}} + \ket{1}_{\mathrm{A}}\ket{0}_{\mathrm{B}} \right). \label{psiplus}
\end{align} 
\end{subequations}
To date, experiments with biphotons have observed one or the other of these Bell state forms; we will show a straightforward method of producing both of these without the need for post-processing of the downconverted photons.

\begin{figure} [h!]
\centering
\includegraphics[width=.99\linewidth]{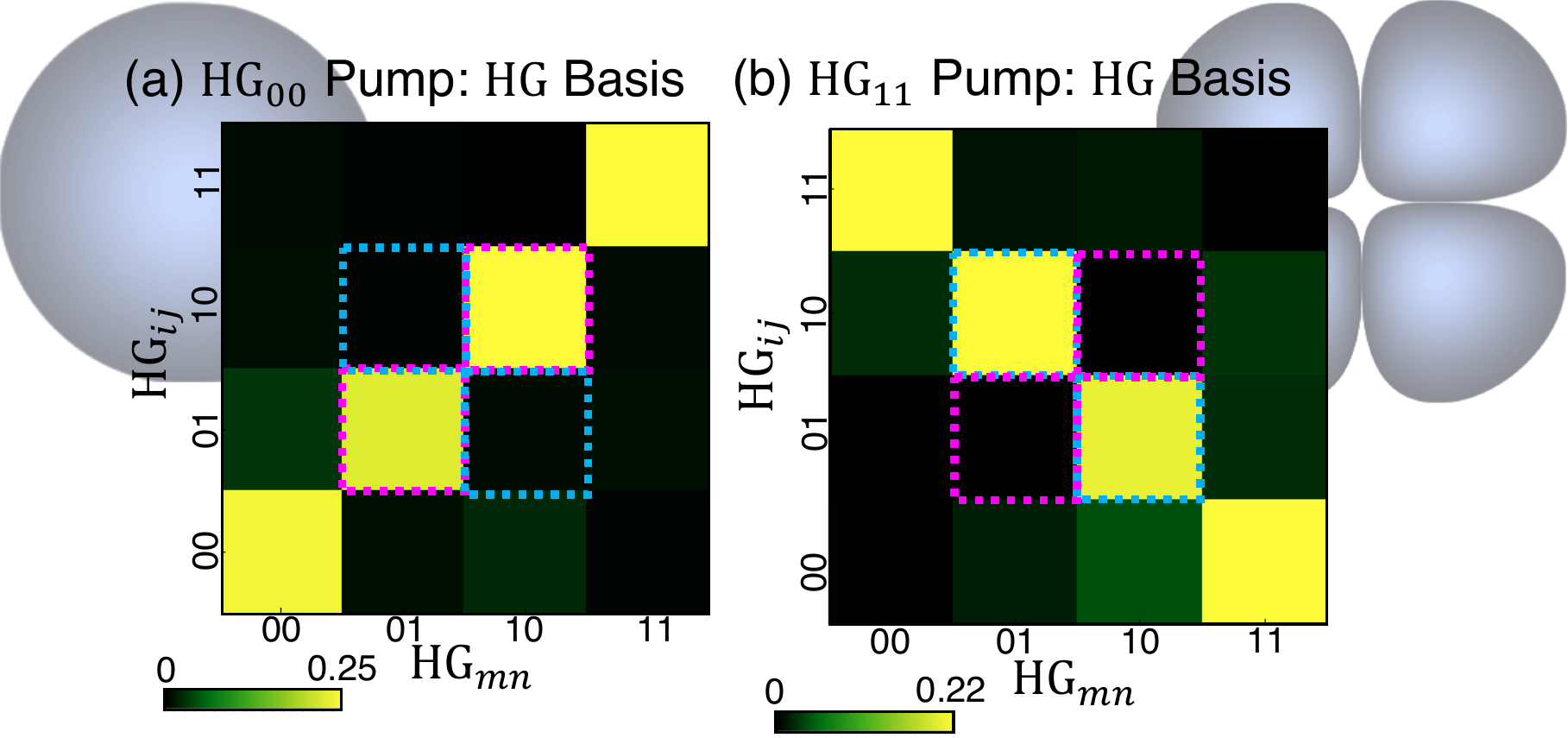}
\caption{(a)-(b) Coincidences measured in the HG basis for biphotons generated by a Gaussian pump beam and an $\ket{\mathrm{HG}_{11}}$ pump beam, respectively. Magenta squares highlight the Bell state form $\ket{\Phi^+}$ (Eqn. \ref{phiplus}), while blue squares highlight the $\ket{\Psi^+}$ form (Eqn. \ref{psiplus}). }
\label{fig:hgspectra}
\end{figure}

The maximally-entangled photon pair generated from a Gaussian pump beam, incident on a Type-I beta barium-borate (BBO) crystal, can be represented as an infinite superposition of either Laguerre-Gaussian (LG) \cite{Salakhutdinov2012Full-fieldPhotons, Restuccia2016ComparingSystem,Valencia2021EntangledEntanglement} or Hermite-Gaussian (HG) \cite{Walborn2005ConservationDown-conversion, Ren2005EntanglementPhotons, Walborn2012GeneralizedConversion, Restuccia2016ComparingSystem,Kovlakov2017SpatialPostselection} modes. Fig. \ref{fig:hgspectra} (a) shows a portion of the (experimental) coincidence spectrum in the HG basis. Focusing on first-order modes, we can identify the presence of certain states---as highlighted in magenta---with the notation of Eqn. \ref{phiplus}, taking $\ket{0} = \ket{\mathrm{HG}_{01}}$ and $\ket{1} = \ket{\mathrm{HG}_{10}}$. The state can be written,
\begin{equation} \label{PhiHG}
    \ket{\mathrm{HG}_{01}}_{\mathrm{A}}\ket{\mathrm{HG}_{01}}_{\mathrm{B}} + \ket{\mathrm{HG}_{10}}_{\mathrm{A}}\ket{\mathrm{HG}_{10}}_{\mathrm{B}} \propto \ket{\Phi^{+}_{\mathrm{HG}}},
\end{equation} 
where the subscript in $\ket{\Phi^{+}_{\mathrm{HG}}}$ denotes the HG basis. Previous theoretical work \cite{Walborn2005ConservationDown-conversion} outlined how the state of Eqn. \ref{PhiHG} could be transformed to the second Bell basis state of Eqn. \ref{psiplus} using supplementary optics subsequent to SPDC. However, experimental implementation of this methodology would require additional optical alignment with concomitant photon loss.

In this paper, we demonstrate direct entanglement of HG modes in a $|\Psi^{+} \rangle$ Bell state, Eqn. \ref{psiplus}:
\begin{equation} \label{PsiHG}
\ket{\mathrm{HG}_{01}}_{\mathrm{A}}\ket{\mathrm{HG}_{10}}_{\mathrm{B}} + \ket{\mathrm{HG}_{10}}_{\mathrm{A}}\ket{\mathrm{HG}_{01}}_{\mathrm{B}} \propto \ket{\Psi^{+}_{\mathrm{HG}}}.
\end{equation}
Its presence is identified with blue dashes in Fig. \ref{fig:hgspectra} (b), made possible by structuring the spatial mode of the pump to engineer the desired biphoton correlations. The ability to pump-tailor entangled photons in a ``balanced'' spatial basis like this comes with potential benefits, including that it is experimentally favorable when considering the symmetrical accumulation of different phases (e.g., geometric or Gouy phases) as the modes propagate through optics. Such a symmetric basis may find applications in quantum sensing and imaging measurements with azimuthal sensitivity. Lastly, the methods we outline here can be extended to the construction of other two-qubit states with structured light, potentially including arbitrary phases between modes, while minimizing optical elements and reducing coincidence probabilities for unwanted photon modes.

Spontaneous parametric downconversion produces a two-photon state that can be represented as the superposition of an infinite number of entangled single-photon HG modes. This follows directly by applying the plane-wave methodology of Walborn~\cite{Walborn2010} to a Hermite-Gaussian basis, just as has been carried out for LG modes~\cite{Franke-Arnold2002}. In brief, the quantum theory of parametric downconversion is derived from a classical description of the nonlinear interaction, followed by quantization of the electromagnetic field. A classical description of the pump beam is then adopted, allowing a quantitative relationship to be formulated between the structure of the pump photons and the weighting of the 2-photon HG states exiting the nonlinear crystal:
\begin{equation}\label{spdc1}
\ket{\Psi} = \sum_{i,j,m,n} C^{ijmn} \ket{\mathrm{HG}_{ij}}\ket{\mathrm{HG}_{mn}}.
\end{equation}
The weighting coefficients are
\begin{equation}\label{spdc2}
C^{ijmn} = \int dr_\perp \psi_{\rm pump}(r_\perp)\mathrm{HG}^*_{ij}(r_\perp)\mathrm{HG}^*_{mn}(r_\perp),
\end{equation}
where $r_{\perp}$ is the position vector transverse to the beam exiting the crystal, $*$ indicates complex conjugation, and $\psi_{\rm pump}(r_\perp)$ characterizes the pump structure. If attention is restricted to Bell states of first-order HG modes, then the output can be truncated to just four modes:
\begin{equation}\label{spdc3}
\{ijmn\}\in \biggl\{ \{0101\}, \{0110\}, \{1001\} , \{1010\} \biggr\}.
\end{equation}

Our work focuses the application of this formalism to two types of pump beams. Pump photons with the transverse structure that is Gaussian deliver the following (re-normalized) coefficients:
\begin{equation}\label{spdc4}
C^{0101} = \frac{1}{\sqrt{2}},  \,\, C^{0110} = 0,  \,\, C^{1001} = 0,  \,\, C^{1010} = \frac{1}{\sqrt{2}}.
\end{equation}
These comprise the $\ket{\Phi^+}$ state of Eqn. \ref{phiplus}. On the other hand, using the transverse structure of an $\mathrm{HG}_{11}$ beam in Eqn. \ref{spdc2} produces a distinctly different set of re-normalized coefficients:
\begin{equation}\label{spdc4}
C^{0101} = 0, \,\, C^{0110} = \frac{1}{\sqrt{2}},  \,\, C^{1001} = \frac{1}{\sqrt{2}},  \,\, C^{1010} = 0.
\end{equation}
These comprise the  $\ket{\Psi^+}$ state of Eqn. \ref{psiplus}. This establishes the theoretical methodology for generating either Bell state without post-processing downconverted photons.

\begin{figure*}[h!]
\centering
\includegraphics[width=0.9\linewidth]{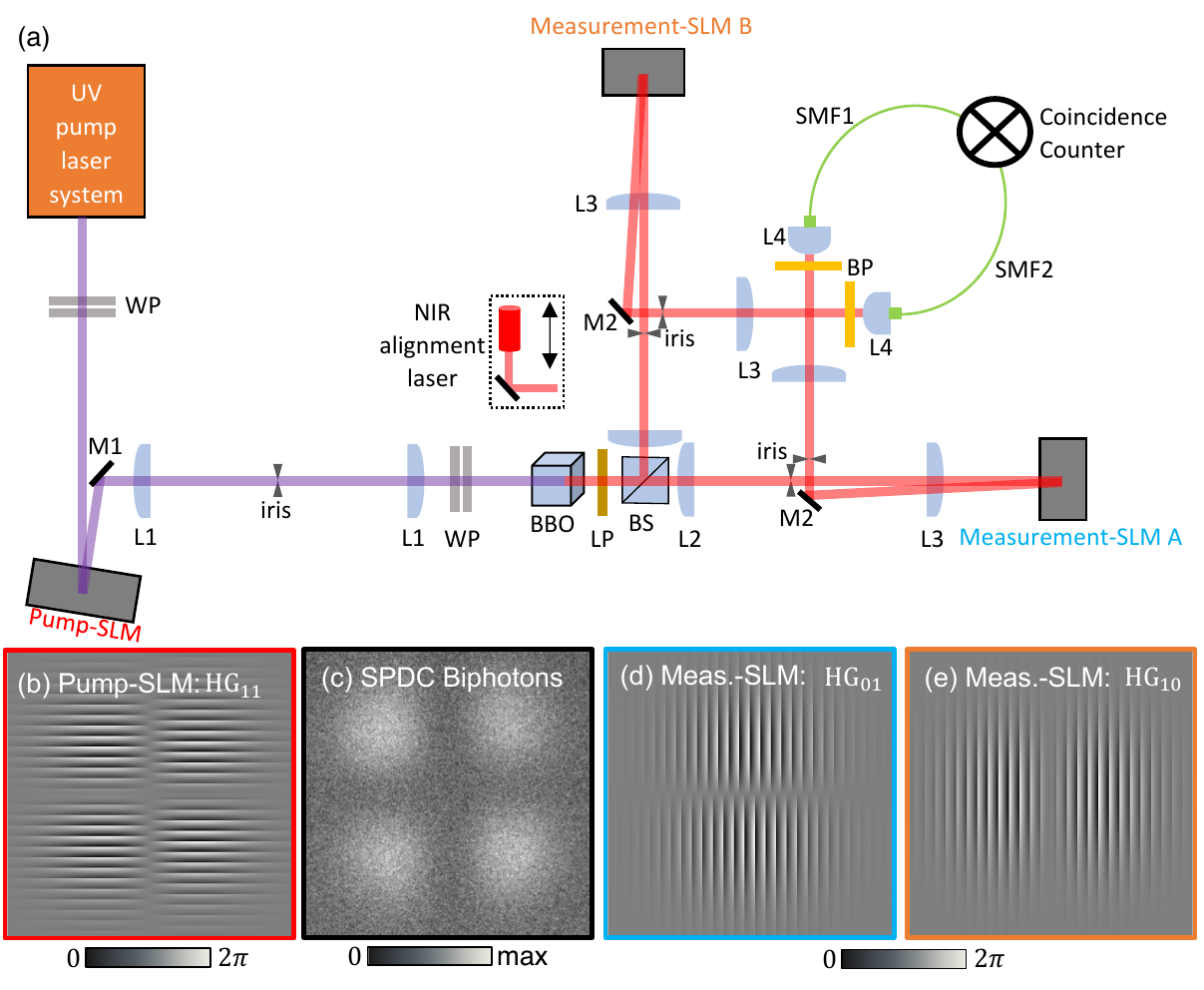}
\caption{(a) Schematic diagram of the experiment. ``UV pump laser system'' includes optical fibers and coupling/collimation optics for generating an initial Gaussian mode. WP: pair of quarter- and half-waveplates; SLM: spatial light modulator; L1: plano-convex lens with $1 \, \mathrm{m}$ focal length; NIR: near-infrared $808 \, \mathrm{nm}$ alignment laser; LP: long-pass filter; BS: $50/50$ beam splitter; L2: plano-convex lens with $10 \, \mathrm{cm}$ focal length; M1: UV-enhanced dielectric mirror; M2: NIR-enhanced dielectric mirror; L3: plano-convex lens with $20 \, \mathrm{cm}$ focal length; BP: band-pass filter; L4: fiber coupling lens with $2.54 \, \mathrm{cm}$ focal length; SMF: single-mode fiber. (b) Representative hologram used on the pump SLM. (c) CCD-captured intensity of the down-converted photons after the BBO. (d)-(e) Representative holograms used on the two measurements SLMs.}
\label{fig:Schematic}
\end{figure*}

We now describe how we experimentally realized the above $|\Psi^{+} \rangle$ state for HG modes. Starting with a Gaussian mode, we sculpted the pump beam to an $\mathrm{HG}_{11}$ mode  with the setup depicted in the schematic of Fig. \ref{fig:Schematic}. The pump beam was a continuous wavelength, ultraviolet (UV, $\lambda = 406$ nm) laser beam; it was coupled into and emerged out of a single-mode optical fiber and was collimated to a waist of $0.87$ mm and then directed towards a phase-only, reflection-based spatial light modulator (SLM)---Holoeye Pluto-2.1-UV-099. On this pump-SLM, a blazed diffraction grating was used as a digital hologram to produce a desired optical mode (i.e., a fully-defined HG mode) in the first-order diffracted beam---Fig. \ref{fig:Schematic} (b) shows an $\mathrm{HG}_{11}$ mode grating. The diffracted beam was subsequently 4f-imaged from the SLM to pump a Type-I BBO crystal, of thickness 5 mm, aligned for collinear SPDC (zero-degree opening angle of the photon cones). The pump power was approximately 35 mW incident on the BBO for a Gaussian pump; for the $\mathrm{HG}_{11}$ pump, the power was approximately 17 mW, where the power loss is due to the amplitude control of the hologram.

After the SPDC process, a longpass filter was placed behind the BBO to block light from the $406$ nm pump beam while allowing transmission of down-converted near-infrared (NIR) photons. The NIR biphotons were then separated into ``signal'' (A) and ``idler'' (B) arms with a 50/50 beamsplitter. In each arm, the output of the BBO was 4f-imaged onto an SLM (Holoeye Pluto-2.1-NIR-118). Fig. \ref{fig:Schematic} (c) shows the 4f-image of the collective biphotons, captured with an Andor iKon-M CCD, that resulted from an $\mathrm{HG}_{11}$ pump profile. The camera was removed for subsequent measurements. Each measurement-SLM displayed a digital hologram of an optical mode to be measured---examples are shown in Fig. \ref{fig:Schematic} (d) and (e). These are functionally correlation filters: if the programmed digital field on each SLM ``overlapped'' with a mode carried by the incident photon in each arm, then a coincidence signal would be detected from successful coupling into the optical fibers. The first-diffracted mode of each measurement-SLM was 4f-imaged onto an apparatus which coupled the result of each applied correlation filter into a single-mode optical fiber. In front of each fiber-coupler was a bandpass filter, $800 \pm 20$ nm, to select the degenerate $812$ nm photons. These fibers were fed into photon counting modules, Excelitas SPCM-AQRH-14-FC. The detectors were connected to an ID Quantique ID900 Time Controller, which measured the coincidence rates seen in the spectra of Fig. \ref{fig:hgspectra}.

With this setup and an $\mathrm{HG}_{11}$ pump, we measured the coincidence spectrum of Fig. \ref{fig:hgspectra} (b). Notice the significant presence of both $\mathrm{HG}_{10}$ and $\mathrm{HG}_{01}$ and the lack of unwanted coincidence counts in $\mathrm{HG}_{00}$ correlation, as opposed to Fig. \ref{fig:hgspectra} (a). For the spectra reported here, coincidences were measured every second for three seconds, and then averaged. These values were normalized so that the sum of the plotted spectrum's coincidence counts is one, as indicated by the plot legends.

\begin{figure}[h!]
\centering
\includegraphics[width=\linewidth]{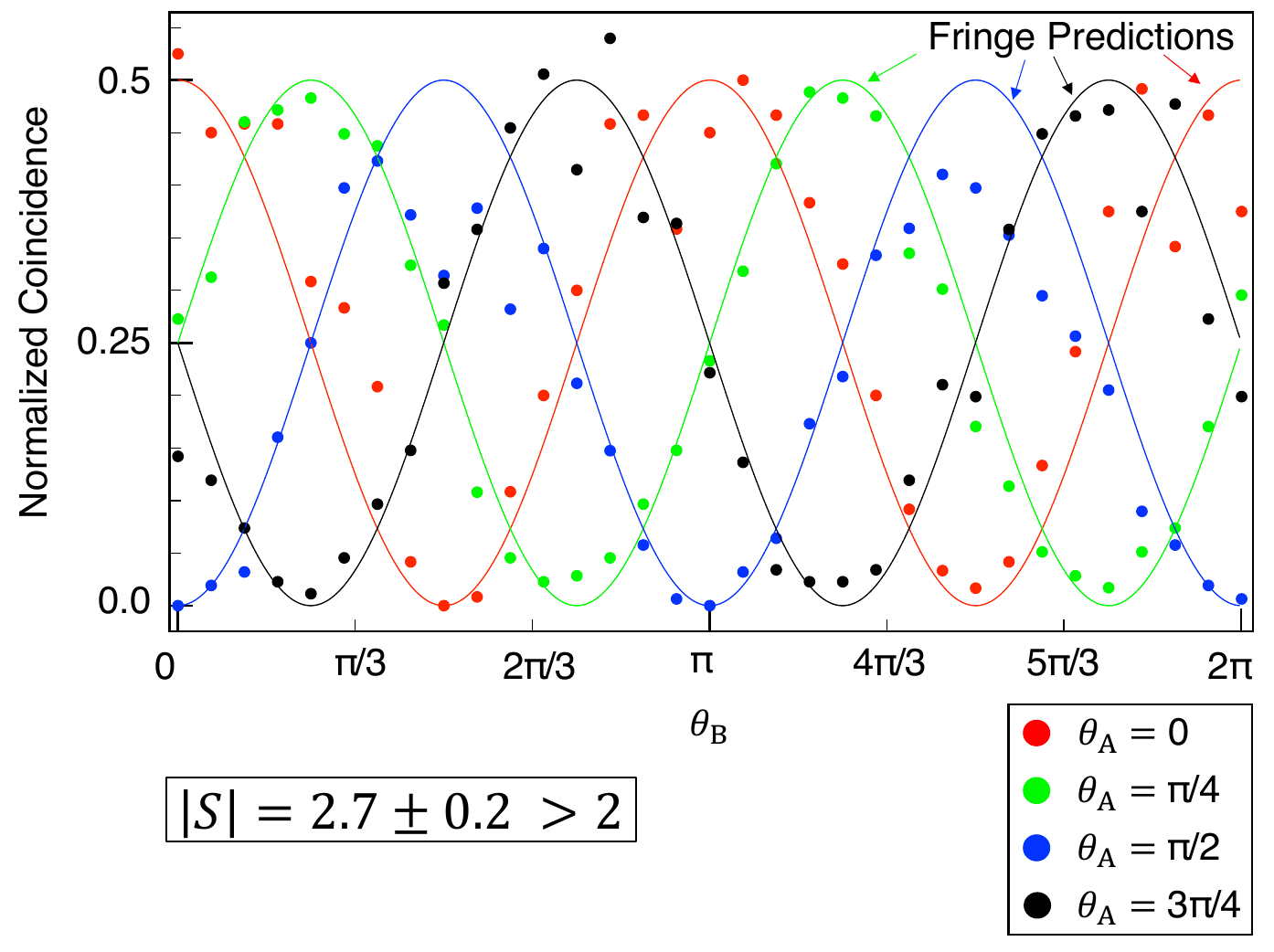}
\caption{Clauser-Horne-Shimony-Holt (CHSH) Bell inequality experimental results to test the state highlighted with blue squares in Fig. \ref{fig:hgspectra} (b).}
\label{fig:chsh}
\end{figure}

In order to test entanglement of the generated HG basis $|\Phi^{+} \rangle$ state, we conducted an experimental test of the Clauser-Horne-Shimony-Holt (CHSH) Bell inequality, adapting the methods of Leach et al. to the HG basis \cite{Leach2009ViolationState-spaces}. This entailed measuring the biphoton state with SLMs that showed holograms of the two superposed modes. Each superposition grating also contained a relative phase shift between the superposed modes; these are $\theta_{\mathrm{A}}$ and $\theta_{\mathrm{B}}$, respectively, and acted as the two test angles for testing the CHSH inequality. These measurement fields had the form:

\begin{align}
    | \theta_{\mathrm{A}} \rangle &= \frac{1}{\sqrt{2}} \left( e^{i \, \theta_{\mathrm{A}}} |\mathrm{HG}_{01} \rangle + e^{-i \, \theta_{\mathrm{A}}} |\mathrm{HG}_{10} \rangle  \right), \\
    | \theta_{\mathrm{B}} \rangle &= \frac{1}{\sqrt{2}} \left( e^{i \, \theta_{\mathrm{B}}} |\mathrm{HG}_{01} \rangle + e^{-i \, \theta_{\mathrm{B}}} |\mathrm{HG}_{10} \rangle  \right).
\end{align}

As shown in Fig. \ref{fig:chsh}, $\theta_{\mathrm{A}}$ was set at four discrete test angles, and for each of these the test angle $\theta_{\mathrm{B}}$ was swept from $0$ to $2\pi$. For each pair of angles, coincidence measurements, $C(\theta_{\mathrm{A}}, \theta_{\mathrm{B}})$, were recorded. The purpose of sweeping the measurement fields like this was to use those coincidence readings to calculate the CHSH-Bell parameter:

\begin{equation}
    S = E(\theta_{\mathrm{A}}, \theta_{\mathrm{B}}) - E(\theta_{\mathrm{A}}, \theta^{'}_{\mathrm{B}}) + E(\theta^{'}_{\mathrm{A}}, \theta_{\mathrm{B}}) + E(\theta^{'}_{\mathrm{A}}, \theta^{'}_{\mathrm{B}}).
\end{equation}
If $|S| > 2$, then the CHSH-Bell inequality is violated and the experimental test cannot be explained by 
classical ``hidden variable'' theories. If violated, the test is a demonstration of the entangled nature of the measured state. Each quantity of $E$ was picked out from the coincidence data at the following particular orientations:
\begin{equation}
\begin{split}
    E(\theta_{\mathrm{A}}, \theta_{\mathrm{B}}) = \frac{1}{D} \left[\right.
    C(\theta_{\mathrm{A}}, \theta_{\mathrm{B}}) + C(\theta_{\mathrm{A}} + \frac{\pi}{2}, \theta_{\mathrm{B}} + \frac{\pi}{2}) -& \\
    C(\theta_{\mathrm{A}} + \frac{\pi}{2}, \theta_{\mathrm{B}}) - C(\theta_{\mathrm{A}}, \theta_{\mathrm{B}} + \frac{\pi}{2}) &
    \left.\right],
\end{split}
\end{equation}
for normalization value
\begin{equation}
\begin{split}
    D = C(\theta_{\mathrm{A}}, \theta_{\mathrm{B}}) + C(\theta_{\mathrm{A}} + \frac{\pi}{2}, \theta_{\mathrm{B}} + \frac{\pi}{2}) +& \\
    C(\theta_{\mathrm{A}} + \frac{\pi}{2}, \theta_{\mathrm{B}}) + C(\theta_{\mathrm{A}} &, \theta_{\mathrm{B}} + \frac{\pi}{2}).
\end{split}
\end{equation}
The coincidence measurements of Fig. \ref{fig:chsh} were also normalized by this value of D for each test angle $\theta_{\mathrm{A}}$. The solid lines are not fits, but are the predicted trends from:
\begin{equation}
    C(\theta_{\mathrm{A}}, \theta_{\mathrm{B}}) \propto \cos^{2}{(\theta_{\mathrm{A}} - \theta_{\mathrm{B}})}.
\end{equation}
The coincidence measurements follow these trends closely and the CHSH-Bell parameter was calculated at $|S| = 2.7 \pm 0.2$, which violates the inequality, helping to confirm the entangled $|\Psi^{+} \rangle$ state of Eqn. \ref{psiplus} inferred from the measurements of Fig. \ref{fig:hgspectra} (b).

\begin{figure} [h!]
\centering
\includegraphics[width=.99\linewidth]{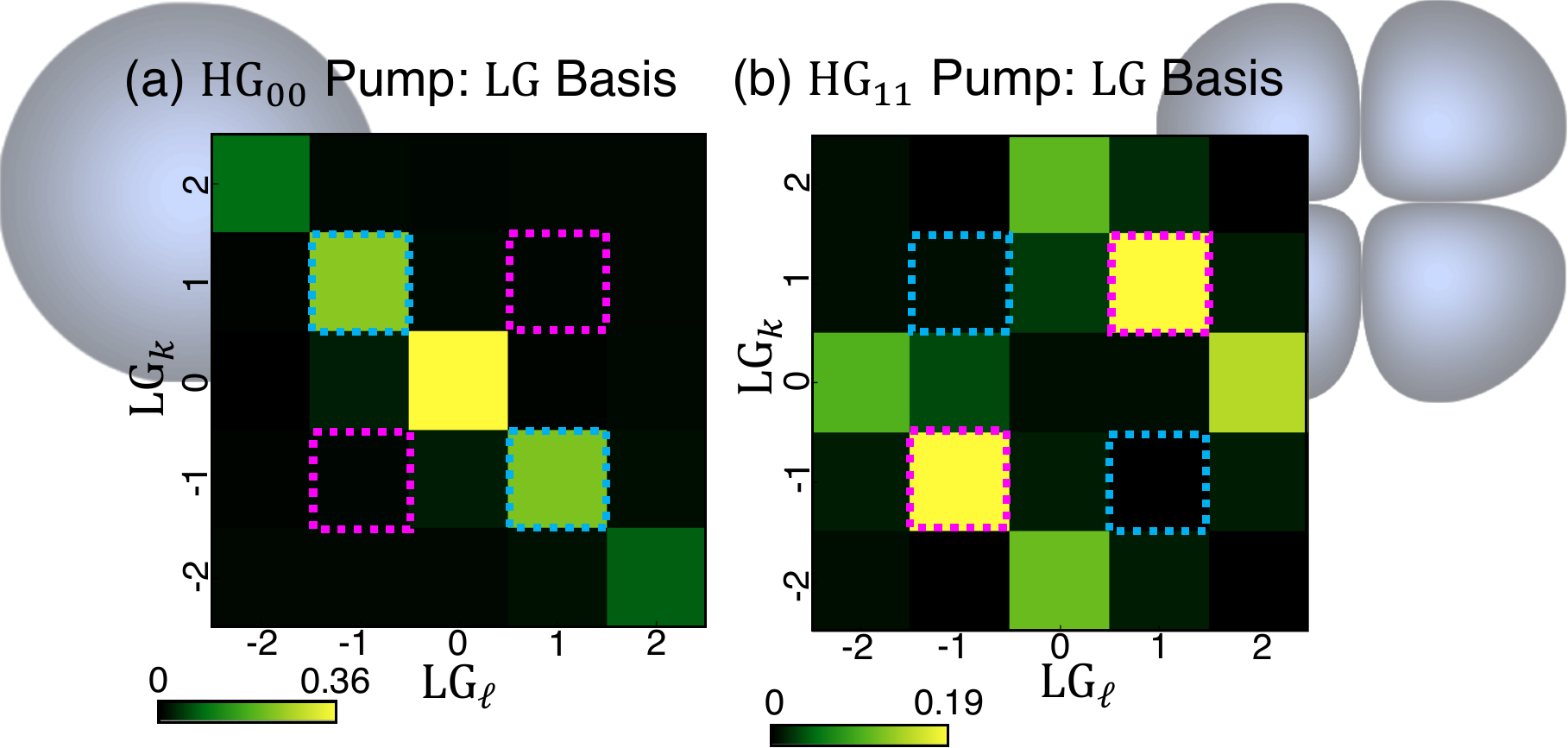}
\caption{(a)-(b) Coincidences measured in the LG basis for biphotons generated by a Gaussian pump beam and an $\ket{\mathrm{HG}_{11}}$ pump beam, respectively. Magenta squares highlight first-order $\ket{\Psi^+}$ states (Eqn. \ref{phiplus}), while blue squares highlight first-order $\ket{\Phi^+}$ states (Eqn. \ref{psiplus}). }
\label{fig:lgspectra}
\end{figure}

Lastly, we note that the complementary basis of Laguerre-Gaussian (LG) modes also shows a change in Bell state representation when an $\mathrm{HG}_{11}$ pump is used. Fig. \ref{fig:lgspectra} (a) shows the measured coincidence spectrum for a Gaussian pump---which, interestingly, has the form of Eqn. \ref{psiplus} (as opposed to the HG basis which was in form of Eqn. \ref{phiplus}). By pumping with $\mathrm{HG}_{11}$, we measured a switch to the Bell state representation Eqn. \ref{phiplus}, as highlighted with magenta dashes in Fig. \ref{fig:lgspectra} (b).

In conclusion, we have outlined and demonstrated a straightforward experimental method for direct generation of the Bell state of mixed qubits ($|\Psi^{+} \rangle$, Eqn. \ref{psiplus}) in the first-order Hermite-Gaussian mode basis of biphotons. The method is easily adaptable: one isolates a desired correlation of biphotons to take the form of a given Bell state (e.g., Eqn. \ref{phiplus} or \ref{psiplus}), and from there search for a pump photon distribution that could yield those correlations. This pump structuring method has the potential to be implemented in experiments to produce all four Bell states, in both pure and mixed forms, a programmable and versatile asset for quantum information science. 

\section*{Acknowledgements}
The authors thank C. Zhu for helpful discussions and P. Fowler, S. Ghazanfarour, and A. Halaoui for experimental support.

\section*{Funding}
W. M. Keck Foundation and National Science Foundation (1553905).

\section*{Disclosures}

The authors declare no conflicts of interest.

\section*{Data availability} 

Data underlying all results presented are available from the authors upon reasonable request.

\bibliographystyle{apsrev4-2}
\bibliography{References}

\end{document}